# The design and performance of the ESS Neutrino Super Beam project ESSνSB

**Tord Ekelof**[1]
*Uppsala University*
*Sweden*
*E-mail:* `Tord.Ekelof@physics.uu.se`

**Marcos Dracos**
*IPHC-IN2P3/CNRS Université de Strasbourg*
*France*
*E-mail:* `marcos.dracos@in2p3.fr`

The European Spallation Source linear proton accelerator will have a uniquely high power of 5 MW. The modifications and additions to the accelerator that are required to enable the generation of a uniquely intense neutrino Super Beam ESSνSB for measurement of leptonic CP violation using a large water Cherenkov detector are described. ESSνSB is complementary to the other two proposed Super Beam experiments Hyper-K and DUNE by the fact that the resulting high neutrino beam intensity allows to place the neutrino detector at the second oscillation maximum. The simulated performance of the ESSνSB experiment is compared to the other two experiments under the assumption of the same systematic errors in all three experiments. The performance of ESSνSB for CP violation precision measurements is found to be highly competitive.

---

[1]Speaker

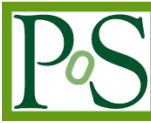

1.      Introduction

The European Spallation Source (ESS) proton linear accelerator (linac) will have an average beam power of 5 MW, which is nearly an order of magnitude higher than any other currently operating proton accelerator in the world. ESS is now well into its construction phase and will deliver first beams in 2021 and be operating at full specification in 2025. An EU supported Design Study of an ESS neutrino Super Beam (ESSvSB) project is being carried out by a European consortium of 17 research groups in 11 countries of using this linac to generate a uniquely intense neutrino beam to be detected in a megaton water Cherenkov neutrino detector located at the second neutrino oscillation maximum where the CP violation signal is ca 3 times larger than at the first oscillation maximum [1].

In section 2 below is described the modifications and additions to the ESS linac the are required to enable an increase of the linac average power to 10 MW and the proposed way to use half of this power to generate a uniquely intense neutrino Super Beam, the other half of the linac power being used to generate spallation neutrons thereby maintaining the original purpose the ESS project. In section 3 is described the performance of ESSvSB for CP violation discovery and precision measurement of the CP phase angle $\delta_{CP}$ as compared with the DUNE and Hyper-K experiments under the assumption of equal systematic errors for the three experiments.

2.      The required modifications and additions to the ESS linac
'

2.1 The Linac

The ESS linac currently under construction will be used to accelerate protons to the energy of 2 GeV in 2.86 ms long 62.5 mA pulses at 14 Hz pulse frequency to be used for spallation neutron production. The low duty cycle of 4% of the ESS linac makes it possible to accelerate additional pulses of H- ions, interleaved with the proton pulses, to be used for the proposed production of a uniquely high-intensity neutrino beam. To increase the average power by 5 MW will require the installation of three new electric substations and new high voltage cables, of 8 more accelerating modules to raise the beam energy from 2.0 GeV to 2.5 GeV, of new klystron collectors to stand the higher average klystron current, of more capacity chargers for the modulators to allow the increase of the pulse frequency to 28 Hz and of an H- sources and the doubling of the very first accelerator components of the linac [2].

2.2 The Target Station

The Neutrino Target Station includes the proton target itself, the hadron collector (magnetic horn), the decay tunnel and the beam dump. The design of a target for neutrino production capable of withstanding the heat load of a 5 MW beam seems not feasible. In order to reduce the heat-load there will be four targets, which will be hit in sequence by the proton pulses using a beam switching system, thereby reducing the beam power hitting each target to 1.25 MW. Following the EUROv studies [3], a packed bed of titanium spheres cooled with cold helium gas has become the baseline design for a Super Beam based on a 4 GeV proton beam with a power of up to 1.3 MW per target.

The current pulse in the hadron collector that focuses in the forward direction the pions produced in the target, before they each decay to a muon and a muon neutrino, will be produced using a pulsed power supply that has been studied in the EUROv project [4]. The very high current (350 kA) pulse required causes a high heat dissipation in the thin walls of the hadron collector. The flat top of the current pulse can for this reason not be longer than the order of a few microseconds. The 2.86 ms length of the pulses from the ESS linac will therefore be compressed by about three orders of magnitude to about 1.3 µs using a ca 400 m circumference accumulator ring.

2.3 The Accumulator Ring

As a first step in the design of this ring, the magnetic lattice of the accumulator ring of the US Spallation Neutron Source (SNS) in Oak Ridge, USA, has been adapted to the higher energy of the ESS beam, using simulations to study different H$^-$ stripping schemes and the accumulator beam stability including space charge effects. With a 2.86 ms pulse length and a 62.5 mA current the number of H$^-$ per pulse will be $11\times10^{14}$. This first study showed that this high number causes problems with stripping-foil heating and space charge effects. The study has therefore been continued assuming that only 1/4 of the total number of particles in one linac pulse is accumulated in the ring. This can be achieved using either 4 stacked accumulator rings, with a beam switching system at injection to distribute the linac beam pulse into the different accumulator rings, or one ring receiving sequentially 4 times more linac pulses, each of a length that is 1/4 of the linac pulse length [5].

In addition is being studied the design for the Near detector on the ESS site and that for the large underground water Cherenkov Far Detector at the location of the second neutrino oscillation maximum ca 500 km away from the ESS site. A design study of a Far Detector MEMPHYS suitable for ESSνSB has been made by EUROν [3].

3. The performance for CP violation discovery of ESSνSB

In Figure 1 is presented the simulated physics performance of the three different proposed long baseline neutrino experiments *assuming the same level of systematic errors of about 3% for all three*. We consider the assumption of 3% systematic error very optimistic but adopt it here in order to enable a comparison on an equal footing with the other two experiments based on published data. The left plot shows the resolution in the measurement of $δ_{CP}$ versus the value of $δ_{CP}$, the middle plot shows the discovery significance for CP violation versus $δ_{CP}$, and the right plot the significance as function of the covered fraction of $δ_{CP}$.

For ESSνSB three cases are shown in Figure 2: two 250 kt detectors in the Garpenberg mine (540 km baseline, blue curves), two 250 kt detectors in the Zinkgruvan mine (360 km baseline, green curves) and one 250 kt detector in the Garpenberg mine and one in the Zinkgruvan mine (black curves). The Hyper-K curve in the middle and right plots and the two resolution values in the left plot for $δ_{CP} = 0$ and $δ_{CP} = π/2$, indicated by the two dotted horizontal lines, are those presented by Hyper-K at the Neutrino 2018 conference. The DUNE curves have been derived using the public GLoBES file released by the DUNE collaboration with its Conceptual Design Report in 2016. Performance predictions for DUNE, assuming 7 years of data taking, were shown by the DUNE collaboration at the Neutrino 2018 conference. For the comparison, in this plot the same simulations were repeated, assuming 10 years of data taking to be in line with the assumptions made for the Hyper-K simulations. The ESSνSB curves have been derived setting the systematic errors to 3% to be in line with the systematic error levels set by DUNE and Hyper-K. The $θ_{13}$ and $θ_{23}$ values for DUNE and ESSνSS have been set to the same values as those used by Hyper-K, again to compare the three experiments on the same footing. These plots clearly demonstrate that ESSνSB has a higher resolution in the measurement of the CP violating angle $δ_{CP}$ and a larger reach for CP violation discovery than the other two experiments.

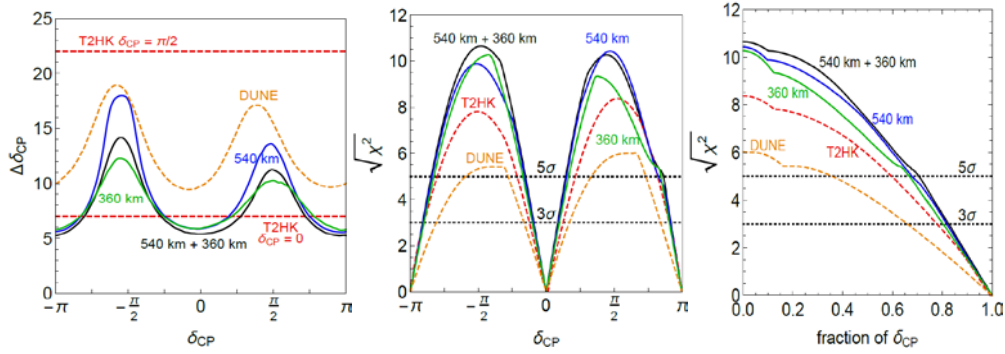

*Figure 1: Resolution in the CP violating angle δ<sub>CP</sub> (left), the CP violation discovery reach in terms of standard deviations versus δ<sub>CP</sub> (middle), and the discovery reach versus the fraction of the δ<sub>CP</sub> range covered (right), for the three different proposed experiments Hyper-K, DUNE and ESSνSB (courtesy E. Fernandez Martinez).*

4. Concluding remarks

The ongoing EU-supported ESSνSB Design Study aims at demonstrating the feasibility of using the uniquely powerful ESS accelerator as proton driver for a neutrino Super Beam experiment. The performance of ESSνSB for CP violation precision measurements is found to be highly competitive with other proposed Super Beam experiments like Hyper-K and DUNE due to the location of the Far Detector at the second neutrino oscillation maximum. An ESSνSB Conceptual Report will be published in 2021.

5. Acknowledgements

The authors are indebted to the participants in ESSνSB for the presentation of the material in this article. The ESSnuSB project is supported by the COST Action CA15139 "Combining forces for a novel European facility for neutrino-antineutrino symmetry-violation discovery" (EuroNuNet). It has also received funding from the European Union's Horizon 2020 research and innovation programme under grant agreement No 777419.